# Ag/Au coated inverted nanopyramids as flexible and wearable SERS substrates for biomolecular sensing

*Anindita Das[a], Udit Pant[b], Cuong Cao[b,c], Rakesh S. Moirangthem[a]\*, Hitesh Bhanudas Kamble[d]*

[a] *Nanophotonics Lab, Department of Physics, Indian Institute of Technology (Indian School of Mines), Dhanbad-826004, JH, India.*

[b] *Institute for Global Food Security, School of Biological Sciences, Queen's University of Belfast, 19 Chlorine Gardens, Belfast, BT9 5DL, U.K.*

[c] *Material and Advanced Technologies for Healthcare, Queen's University of Belfast, 18-30 Malone Road, Belfast, BT9 5BN, U.K.*

[d] *IITB Nanofabrication Facility, Department of Electrical Engineering, Indian Institute of Technology, Bombay, India.*

**Abstract**

Surface enhanced Raman spectroscopy (SERS) has established itself as a promising tool in optical sensing technology. Efforts have been made to improve practicalities of the technology with regards to costs of production, stability, reproducibility, flexibility and robustness. Here, we demonstrate a method to fabricate Ag/Au bimetallic inverted nanopyramid (i-NPyr) and upright nanopyramid (u-NPyr) using a multi-step molding process on flexible plastics as SERS sensors for the ultrasensitive detection of haemoglobin at a very low concentration (down to 1 nM). First, a Si i-NPyr master was created using electron beam lithography, then i-NPyr and u-NPyr structures were imprinted on flexible polymer substrates using nanoimprinting lithography, and last, Ag/Au coatings were coated on top of them. The SERS activity of the imprinted i-NPyr and u-NPyr substrates were evaluated using rhodamine 6G (Rh6G) as probe molecules. Enhancement factor (EF) values of $3.88 \times 10^6$ and $7.86 \times 10^5$ respectively for the i-NPyr and u-NPyr substrates, and the i-NPyr SERS substrate was able to detect Rh6G at concentrations as low as 1pM. In addition, we investigated the stability and resilience of i-NPyr by thoroughly analysing signal performance under angled bending and delicate crumpling. Finally, a proof-of-concept application as a wearable i-NPyr SERS sensor was demonstrated by detecting the analyte in sweat excreted during running and walking. We believe that this extremely sensitive i-NPyr SERS substrate with dense electric field confinement around hotspots (confirmed using computational simulations), as well as its good stability, durability, and low production cost, may enable in-situ measurement of analytes in wearable technologies.

# 1. Introduction

Using portable medical devices, and monitoring of multiple disease biomarkers has drawn massive interest due to enabling faster, more sensitive, accurate, and real-time diagnosis. The rational design of flexible and wearable sensors has made substantial advances in personalized healthcare solutions [1][2][3]. Unlike rigid counterparts, flexible sensors are soft and can conformably attach to bio-interfaces of random shapes (such as oral, eye and epidermis). The wearable sensors are mainly used to track the wearer's motion and vital signs (heart rate, blood pressure and oxygen saturation levels, etc.) [4]. However, in-situ chemical analysis of biofluids (sweat, urine, and saliva, tear) excreted from wearer's body is crucial to understand different pathological and physiological conditions [5]. Further sweat sample can provides information regarding multiple analytes such as ammonia, lactate, ethanol, urea simultaneously. Thus, wearable sweat sensors can measures many different constituent molecules and could support the non-invasive detection, prevention and prognosis of diseases [6]. Electrochemical sensor is one of the most popular wearable sensing techniques enabling rapid and sensitive detection with good biocompatibility. However, their complex fabrication process and low selectivity in simultaneous testing of multiple biomarkers are some major challenges [7][8]. Surface-enhanced Raman scattering (SERS) spectroscopy-based detection is considered to be an excellent label-free analytical method for trace molecules, proteins structural characterization and quantification, with sensitivity down to single-molecule [9][10]. As per previous reports, the use of SERS technique significantly improves the applicability of Raman spectroscopy in haematological analysis such as detection of anemia, leukemia, nanotoxicity in cells, glucose level monitoring and cancer screening [11][12][13][14][15]. Further, wearable SERS biosensors hold considerable promise for multiplexed chemical analysis of bio-analytes present in biofluids. Several studies are evidencing the SERS based sensors's non-invasive sensitive sensing ability with high specificity [16][17]. Wu J. et al.

investigated the multiplex detection of three different hepatocellular carcinoma-related microRNA biomarkers using fractal Au nanotags [18]. But their mass production with minimal cost and uniform structural reproducibility remains challenging due to their complex fabrication protocols. Hence, fabrication of flexible, sensitive, bio-compatible, stable, and cost-effective SERS substrate has been in high demand.

Adequate number of reports suggested that lithographically patterned homogeneous nanostructure arrays with ultrasmall nanogaps or periodicity can utilize the electromagnetic (EM) interactions between adjacent nanostructures and enhance the Raman signal significantly [19]. Despite the fact that both electromagnetic (EM) and chemical mechanisms (CM) are responsible for SERS performance, the EM contribution due to surface plasmon resonances from rough metallic surfaces is much larger and predominate [20]. Therefore, advanced nanostructures with sharp edges, and nanoscale tips can generate large number of SERS-active electromagnetic hotspots due to lightning rod effect, leading to better SERS enhancement [21][22]. The commercially available textured Si substrate, commonly known as Si Klarite substrate has been shown to have great SERS activity that originates from its inverted pyramid structures. Both straight and inverted nanopyramid substrates have already employed as sensor chip for the SERS-based detection of atmospheric bioaerosols [23], explosive chemical traces [24], viruses [25], microplastics [26]. However, the high processing cost of commercial Si inverted nanopyramid (i-NPyr) or upright nanopyramid (u-NPyr) nanopatterned standard substrates limits their mass fabrication and large-scale practical application. To reduce fabrication costs without sacrificing quality, researchers have started using nanoimprint lithography to create multiple copies of nanopatterned Si master stamps on flexible polymer or rigid substrates with high structural reproducibility and low processing costs [20][27][28][29].

It has been reported that silver nanostructures with small interparticle gaps can have strong SERS enhancement factor (EF) due to the interband transition in the UV region being well separated from its plasmon band, causing lesser plasmon damping. In contrast, gold has both interband transition and plasmon band in the same spectral region [30]. However, compared to gold silver has lesser biocompatibility and stability as it degrades through oxidation in the air over time [31][32]. Therefore, designing hybrid composite flexible SERS substrate is needed to improve the silver-based SERS devices' lifetime. Here, we present flexible plastic substrates patterned with inverted (i-NPyr) and upright nanopyramids (u-NPyr) with Ag/Au bimetallic coatings for label-free SERS based bio-sensing applications. With ease in nano-patterning directly onto polymer films with nanoimprinting lithography (NIL), the developed integrated fabrication using electron beam lithography (EBL) and multistep replica moulding gives significant progress in making cost-effective i-NPyr, and u-NPyr patterned flexible SERS platforms. After evaluating the SERS signalling ability using Rh6G as a probe molecule for both the designs, the i-NPyr structures with improved EF were chosen to be investigated further to prove its potential as a wearable biosensor. Since SERS enhancement depends on the excitation angle and distribution of plasmonic hotspots, the signal may be greatly unstable under the mechanical deformations of the flexible sensors. Hence, SERS sensor reliability and stability under a wide range of excitation angles and mechanical strains which may arise under body movements is a tremendous challenge. This work expresses our efforts to examine the SERS vibrational spectrum of hemoglobin (Hb) protein molecules, the most common type of cellular component in blood and its detection at low concentrations using i-NPyr SERS substrate. Further, to show the practical utility of our fabricated i-NPyr substrates, we have tested the sensor's biosensing applicability under different mechanical conditions by detecting Hb proteins and monitoring human sweat sample. EM SERS enhancement was further analysed from a theoretical point using finite element method (FEM) based simulations.

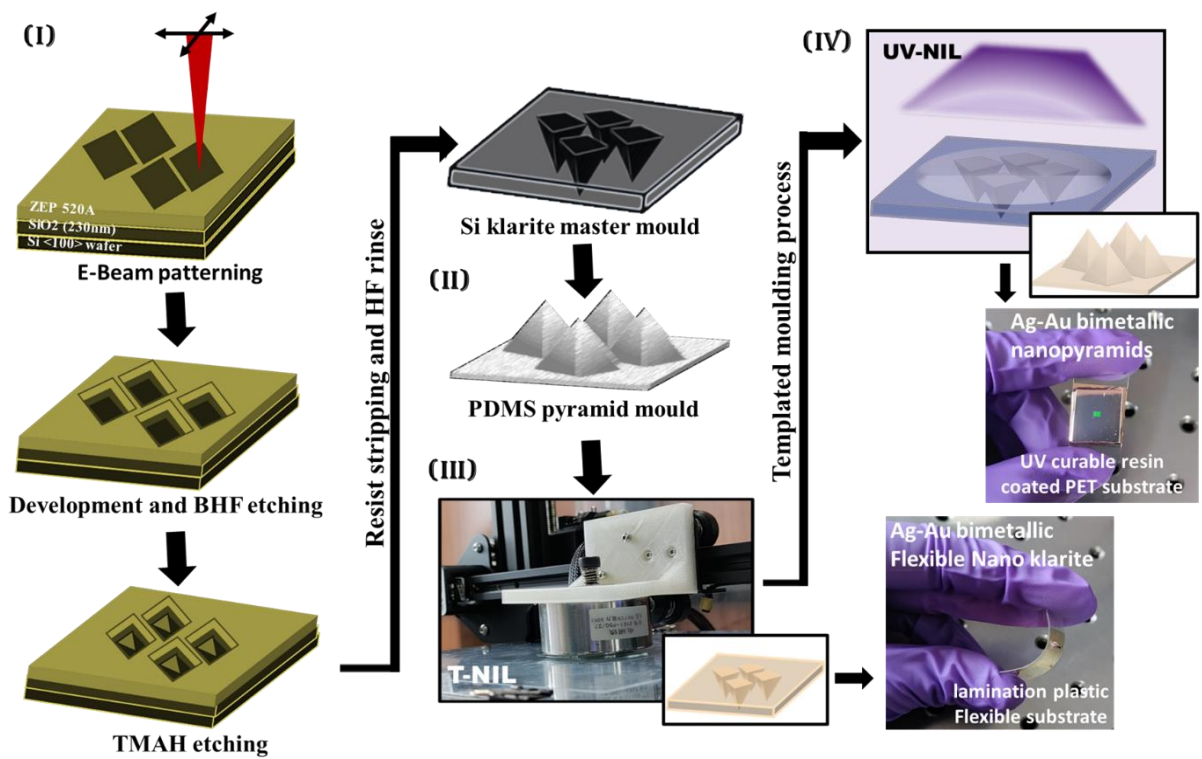

**Figure 1.** Schematic of the fabrication route of Ag/Au bilayer i-NPyr and u-NPyr array as flexible SERS substrates.

## 2. Experimental Section

### 2.2 Fabrication of Si i-NPyr

By following our previously reported fabrication protocol with simple variation in structural geometry and periodicity, Si i-NPyr master mould was fabricated (Figure 1 (I)) [33]. In details, the silicon wafer was decontaminated through famous Radio Corporation of America (RCA) cleaning process [34] and then 230 nm thick silicon oxide ($SiO_2$) layer was deposited using a custom-built wet oxidation furnace. The ZEP 520A photoresist of thickness 300 nm was spun onto the processed wafer and baked for 10 min at 180 °C. The EBL writing was done at 20 kV acceleration voltage with 10 µm aperture size and 10 mm working distance. The e-beam with a dose factor of 50 μC/cm$^2$ was used to expose the resist to create nano square (1250×1250 nm) arrays with zick zack periodicity over 1.5 mm$^2$ scan area. After writing, exposed resist parts were developed for 5-6 min using ZED-N50 developer solution followed by 1-minute plasma ashing with 40 sccm O$_2$ gas flow rate at 100 watts power to eliminate the residual resists from the developed parts. Further, a post-exposure bake for 1 min at 90 °C could improve the adhesion of the unexposed resist. Next, for the isotropic etching of the SiO$_2$ layer, the developed wafer was dipped inside a buffered hydrofluoric acid (5:1 BHF) chemical etchant for 2 min and then exposed the Si of the patterned area for further process. The anisotropic etching of the Si from the patterned area was done using 25% aqueous tetramethyl ammonium (TMAH) solution at 90 °C for 1 min 45 s. The Si i-NPyr substrate was fabricated by anisotropic etching of the (111) plane.

### 2.3. Fabrication of Ag/Au bilayer i-NPyr and u-NPyr array

The idea of making polymer SERS substrate through multistep moulding is depicted Figure 1 (II-IV). In the PDMS replication step, the uncured PDMS (a mixture of sylgard polymer base with the curing agent at a 10:1 ratio) was poured on the patterned wafer and degassed then cured at 80 °C for 2 hrs. Prior to PDMS replications, the Si i-NPyr substrate was modified using TMCS chemical in the gas phase silanization process to increase the hydrophobicity of the surface [35]. This is important for the PDMS replication process to be successful as the hydrophobic surface releases the cured PDMS polymer entirely without any permanent damage to the master mold by residue polymers inside the nanopatterns. The resulted cured PDMS polymer with nanopyramidal arrays substrate then subjected to the NIL-based replica molding process.

The flexible plastic i-NPyr substrates were fabricated using our customized T-NIL setup. The PDMS master was pressed over a lamination plastic substrate at 80 °C for 20 min. The imprinting temperature was maintained higher than the PEVA polymer melting temperature (≈ 65 °C) to transfer the nanopattern from PDMS to the molten substrate under a certain pressure applied using electromagnetic force as shown in Figure 1 (III). After cooling, the PDMS mold was carefully peeled off and u-NPyr patterns were replicated as i-NPyr on lamination plastic. Because of the simple operation of T-NIL, it is easy to fabricate multiple SERS substrates with low processing cost and time. In another approach, using UV-NIL technology multiple nanopyramidal substrates were fabricated using lamination plastic i-NPyr substrate as molding stamp. The coated UV polymer resin was cured under UV source (380 nm) for 4 to 5 min, followed by slowly peeling off the lamination plastic from the cured resin layer. Finally, in vacuum metallization 45 nm of Ag and 5 nm of Au were deposited over lamination plastic i-NPyr and UV curable resin u-NPyr substrates using e-beam evaporation and the deposition rate was 1 nm/s. The fully fabricated substrate flexibility is shown in Figure 1. Both types of plasmonic substrates were used to evaluate their efficacy as SERS substrate.

**2.4. Substrate preparation for SERS based detection**

The SERS performance of both type of substrates was evaluated using Rh6G molecules. The SERS spectra for different Rh6G concentrations (from $10^{-6}$ M to $10^{-12}$ M) were collected by drop-casting and drying 0.5 µL of Rh6G solutions on Ag/Au i-NPyr substrates and u-NPyr substrates. A 532 nm diode laser with 0.1 mW power was used as excitation source to acquire all SERS spectra over a 5 s of acquisition time. The total number of acquisitions was 3. A 100× objective lens and this yield the laser spot of area 0.4082 µm$^2$ on the SERS substrates. Also, to ensure reproducibility and signal uniformity, signals were recorded from multiple locations. Further, SERS mapping images of Rh6G molecules were recorded over an area of (X × Y) = (8 × 7) µm$^2$ with a point mapping step of 1 µm. Also, the normal Raman spectrum of the Rh6G solution at $10^{-3}$ M concentration was collected by casting a drop of 0.5 µL on bare Si substrate as a control. The fabricated i-NPyr substrate were further used for the detection of Hb protein molecule. The details of the average EF calculation are provided in the supplementary material.

Haemoglobin solutions of concentration ranging from $10^{-4}$ to $10^{-9}$ M were prepared in deionized water. The SERS measurements were done by drop casting and drying a drop of 0.5 µL for each concentration on i-NPyr substrates. Raman spectra of the lyophilized protein powder and in solution ($10^{-4}$ M) form were also measured on bare Si substrate as reference. To ensure signal homogeneity and reproducibility, SERS spectra were collected from different location of the substrate. Further SERS signal stability under different mechanical conditions was also tested by substrate bending and multiple crumpling tests. Later the fabricated sensors wearability and sweat monitoring ability were also demonstrated by adhering the fabricated i-NPyr sensor to the volunteer's neck under three separate physical conditions (sitting, walking, and running). During sweat monitoring, the observed testing zone temperature was ≈ 25-30 °C, with a humidity of 64%.

## 3. Results

### 3.1 Structural characterization of Ag/Au bilayer i-NPyr and u-NPyr array

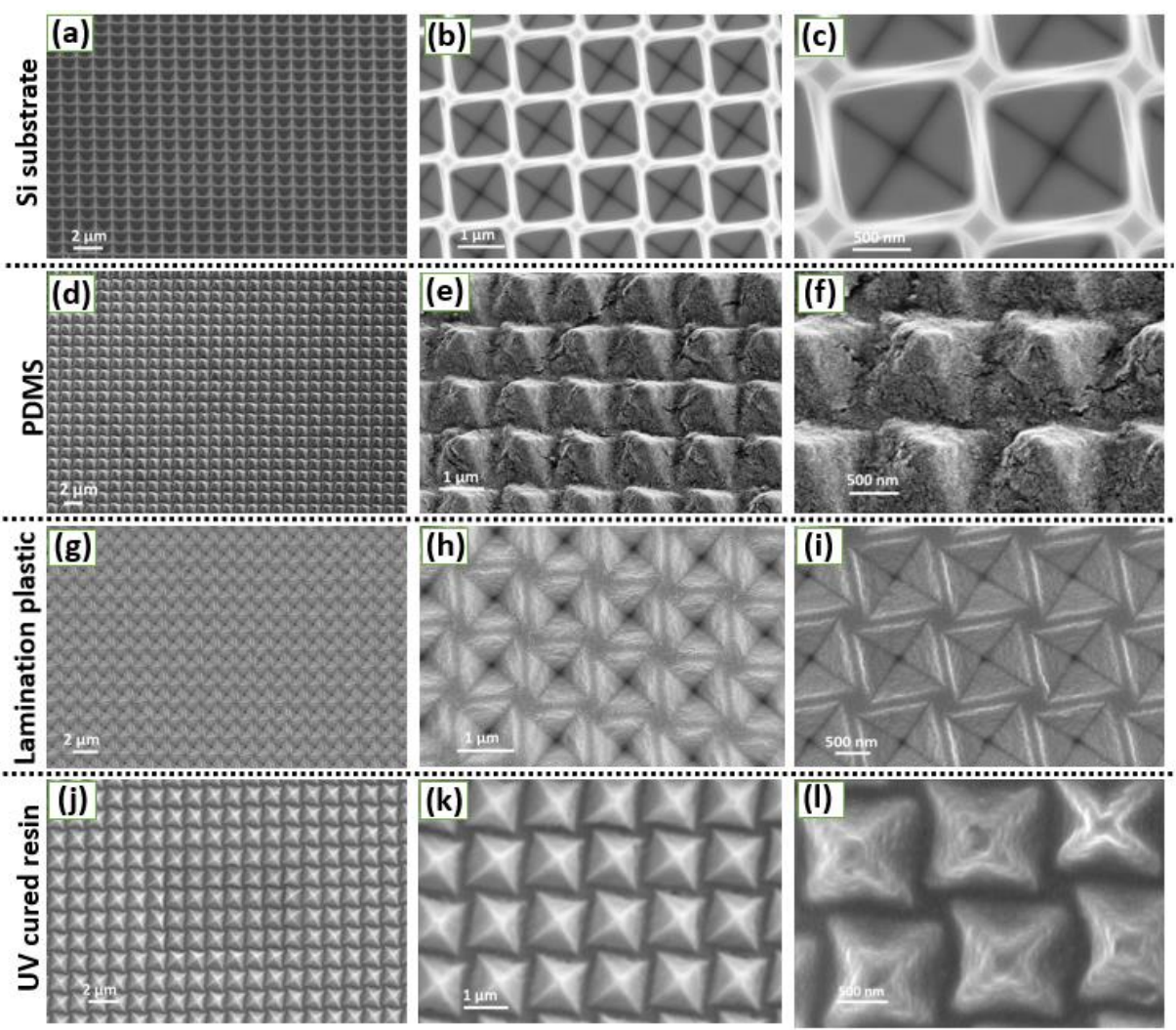

**Figure 2.** SEM micrographs of fabricated nanostructures through EBL and sequential replica moulding. **(a)** under 36° sample tilt and **(b-c)** the top view of the EBL fabricated Si i-NPyr substrate, **(d-f)** images of PDMS u-NPyr mould after replication under 40° sample tilt, **(g-i)** top view of lamination plastic i-NPyr SERS substrate, and **(j-l)** top view of the U.V. resin u-NPyr SERS substrate.

The SEM images in Figure 2 show the nanostructures pattern obtained in four fabrication pathways, including intermediate moulding steps. Figure 2 (a) shows a well order and perfectly fabricated i-NPyr arrays over a large area and the structures are well shaped with sharp tips. The measured single Si i-NPyr tip diameter is 16.58 nm with an apex angle of 59.5° shown in the inset of Figure S1 (In supporting information). Thus, our proposed optimised recipe can be used to generate uniform u-NPyr arrays over Si substrate. The fabricated PDMS pyramid replica from the Si i-NPyr substrate is shown in Figure 2 (d-f), which confirms the successful replication of nanostructures from Si i-NPyr. The replica structure was used as a master mould for the rest of the fabrication. With this way, Si template can be used only once at the beginning and does not involve in rest of the multistep sequential moulding process. The observed surface cracks and damages on the PDMS surface were caused by the e-beam during imaging. The surface morphology of the NIL fabricated i-NPyrs and u-NPyr are shown in Figure 2 (g-i) and 2 (j-l) respectively.

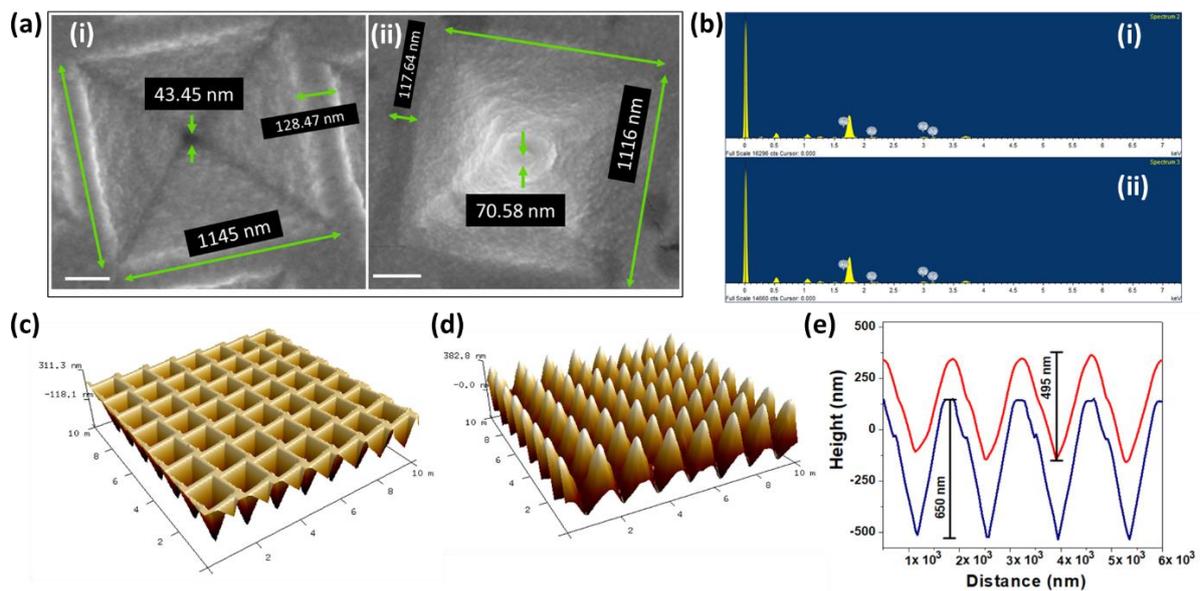

**Figure 3. (a)** The magnified FESEM images of Ag/Au coated single polymer (i) i-NPyr, and (ii) u-NPyr nanostructures showing geometrical details analysed via ImageJ software, scale bar: 200 nm, **(b)** showed their corresponding EDX spectra. **(c-d)** 3D AFM topographical images of both Ag/Au coated polymer i-NPyr and u-NPyr substrates and **(e)** corresponding line profiles (blue; i-NPyr and red; u-NPyr) respectively.

The geometrical information of fabricated polymeric nanostructures are shown in Figure 3 (a). The u-NPyr (Figure 3 (a-i)) and u-NPyr (Figure 3 (a-ii)) have nano tips with measured diameters are 43.45 nm and 70.58 nm respectively. The gap between two adjacent pyramidal pits was around 128.47 nm with square base length of 1145 nm and after replication the gap was around 117.64 nm between two adjacent straight pyramids and has square base length around 1116 nm. The elemental composition of the deposited Ag/Au films on polymer substrates were analysed through SEM-EDX analysis. The measured EDX spectra of i-NPyr substrate is shown in Figure 3 (b-i). and the calculated weight and atomic percentage of Ag, Au are 67.35 %, 32.65 % and 79.02 %, 20.98 % respectively. The measured EDX spectra of u-NPyr substrate is shown in Figure 3 (b-ii). and the calculated weight and atomic percentage of Ag, Au are 67.93 %, 32.07 % and 79.45 %, 20.55 % respectively. The corresponding EDX spectra is shown in Figure 3 (b-i and -ii), where the characteristic peaks associated to Ag and Au elements were detected. As we see that, both the SERS substrates, the thin Au layer covered uniformly over the Ag film. The deposited ultra-thin gold layer not only protects the silver from oxidation but also provides a hybrid plasmonic effect and enhances the SERS activity. The surface analysis part also includes the AFM results, which enables the wedge profile, depth, and height of both type of SERS substrates. Figure 3 (c) and (d) reports the AFM topography of i-NPyr and u-NPyr substrates. The line profiles of both kind of substrate topography are drawn in Figure 3 (e) and shows that the measured depth was found to be around 650 nm for the i-NPyr structures, and the measured height was found to be around 495 nm for the u-NPyr. Hence, both SEM and AFM analysis provide the evidence regarding the regularity and uniformity of u-NPyr and i-NPyr structures over the entire patterned area.

### 3.2 SERS based detection

**3.2.1 SERS effect evaluation.** To report the fabricated Ag/Au coated i-NPyr and u-NPyr structures as SERS substrate, we recorded Raman signal from the substate using 532 nm laser as excitation source. Figure 4 (a) shows the Raman spectra (green) acquired for $10^{-3}$ M Rd6G dye on the bare Si substrate and SERS spectra of $10^{-6}$ M Rh6G dye on Ag/Au nano i-NPyr (blue) and u-NPyr (pink) substrates. The characteristic Raman peaks of Rh6G dye molecules are significantly enhanced for nano i-NPyr and u-NPyr substrates compared to those on the bare Si substrate. Again, to measure the Raman signal enhancement ability of the fabricated i-

NPyr and u-NPyr substrates, a comparative study based on their SERS response using Rh6G as a probe molecule is summarized in Figure 4 (b-f). The SERS spectra of Rh6G for various concentration ranging from $10^{-6}$ to $10^{-9}$ M on both the i-NPyr and u-NPyr substrates are shown in Figure 4(b) and (c) respectively.

The qualitative SERS evaluation of Rh6G molecule for both the substrates is shown in the calibration curves plotted between SERS signal intensity (for peak 612 cm$^{-1}$) vs logarithmic concentration as presented in Figure 4 (d). The calibrated regression equation for the i-NPyr substrate is given by: $I_{612} = 7291.9 \text{ Log } C_{Rh6G} + 68628.2$ with $R^2$ of 0.96 and for the u-NPyr substrate is: $I_{612} = 4709.7 \text{Log } C_{Rh6G} + 42851.4$ with $R^2$ of 0.76. Nearly exact linear relationship can be observed for i-NPyr substrates. Further, to determine the detection limit of both kind of substrates, another set of measurements were performed at lower target molecules concentration of $10^{-12}$ M was done and the results are included in Figure 4 (e) and (f). The limit of detection (LOD) at sub-nanomolar concentrations is essential in biomedical applications for reliable detection as most of the biomolecular Raman spectra are rich and variable, and the effects of peripheral substituents and of distortions imposed by the protein are not easy to decode. The fabricated Ag/Au i-NPyr substrate could detect the analyte molecules concentration down to $10^{-12}$ M whereas the Ag/Au u-NPyr SERS substrate could barely detect the target molecule at concentration down to $10^{-10}$ M. The i-NPyr substrate has proven to be capable of detecting Rh6G molecules with weak characteristic peaks at $10^{-6}$ M concentration. At such low concentrations, the less SERS intensity is quite admissible (inset of Figure 4(e)). However, it is undeniable that it shows fingerprint Raman bands of Rh6G since the observed Raman peaks at this concentration has similarity with the peaks present at other higher concentrations, thus supporting the high sensitivity of the i-NPyr substrates.

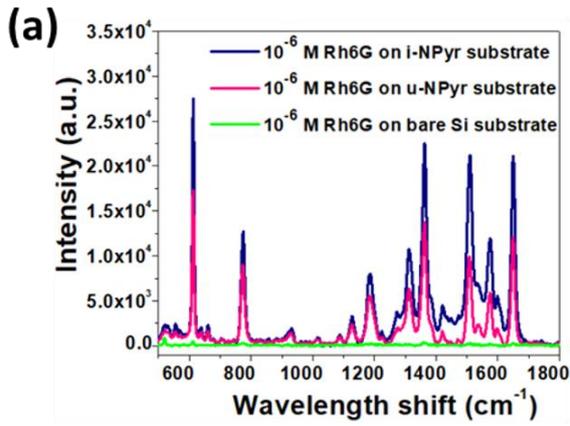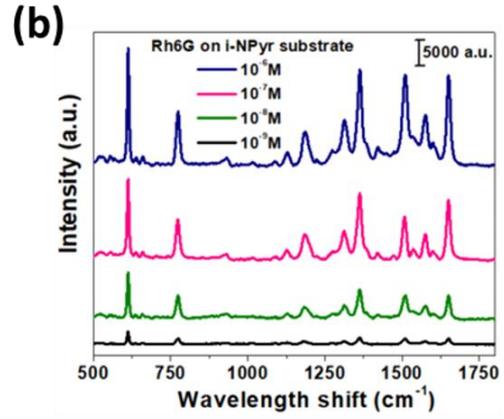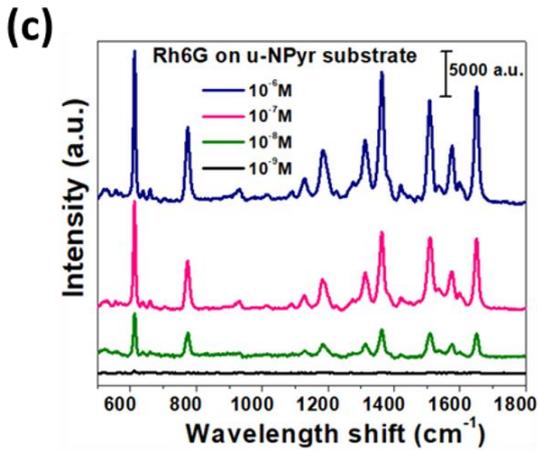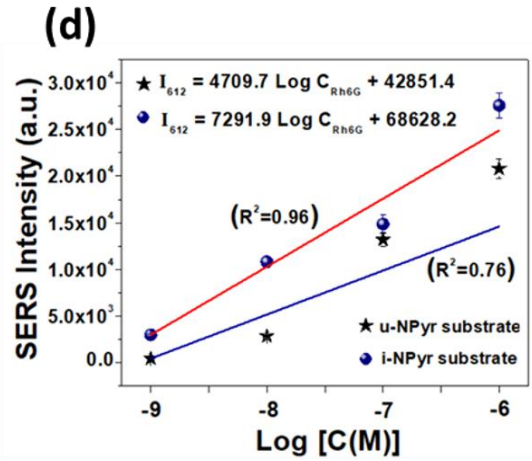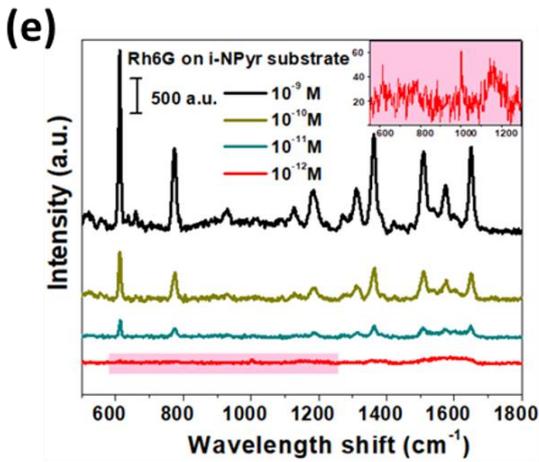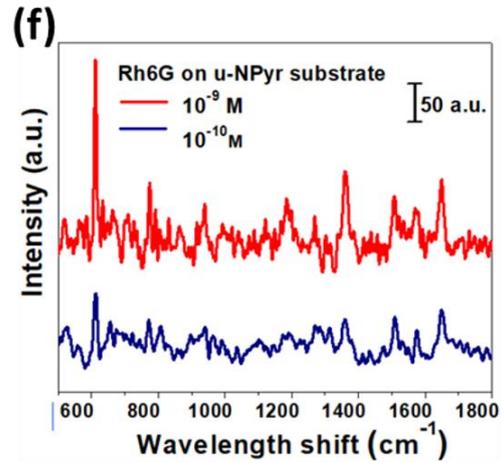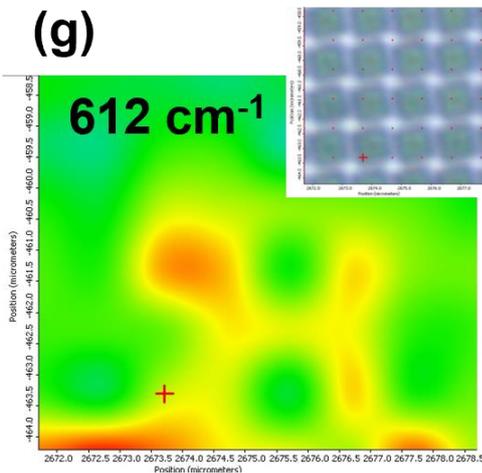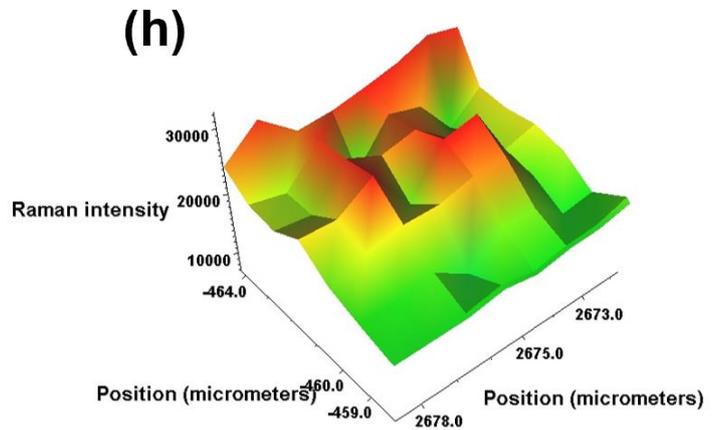

**Figure 4.** SERS spectral analysis of Rh6G molecules. **(a)** Normal Raman spectrum from $10^{-3}$ M Rh6G dye on the bare Si substrate and SERS spectra of $10^{-6}$ M Rh6G dye on Ag/Au nano i-NPyr and u-NPyr substrates. SERS spectra for different concentrations ranging from $10^{-6}$ to $10^{-9}$ M of Rh6G collected from **(b)** Ag/Au nano i-NPyr and **(c)** Ag/Au u-NPyr substrates and the corresponding calibration curve for both type of substrates is shown in **(d).** The Rh6G molecule's limit of detection (LOD) decreased up to **(e)** $10^{-12}$ M concentration in Ag/Au nano i-NPyyr substrate and **(f)** $10^{-10}$ M concentration in Ag/Au u-NPyr substrate. **(g)** optical microscopic image of the mapped area of the Ag/Au nano i-NPyr substrate, both **(h)** 2D, and **(i)** 3D Raman mapping images for the $10^{-6}$ M Rh6G adsorbed Ag/Au i-NPyr substrate for 612 cm$^{-1}$ Raman band.

To test the signal homogeneity, SERS spectra were collected from 10 different locations of each SERS substrate with different analyte concentrations and the results for $10^{-6}$ M analyte concentration are included in Figure S2 on supplementary information. It was found that both the fabricated substrates have good signal reproducibility and homogeneity throughout the surface. Furthermore, the EF of i-NPyr and u-NPyr substrates were calculated by comparing the intensity of the Rh6G major Raman band at 612 cm$^{-1}$ at $10^{-9}$ M concentration. As a reference the measured Raman spectrum of the $10^{-3}$ M Rh6G on bare Si substrate was recorded (see Figure S3). The details of the EF calculations are included in supplementary information section (S3). The calculated EF for i-NPyr and u-NPyr substrates at $10^{-9}$ M Rh6G concentration are 3.88 x $10^6$ and 7.86 x $10^5$ respectively. The i-NPyr substrate has larger SERS EF than the u-NPyr substrate. The Raman intensity mapping over a large area ($8 \times 7$ μm$^2$) of the i-NPyr substrate also included in this study for $10^{-5}$M Rh6G and the bright-field microscopic image of the mapped surface is shown in Figure 4 (g). The 2D Raman map as indicated in Figure 4 (h) shows the SERS hot spots (red colour) for the Rh6G 612 cm$^{-1}$ peak. The corresponding 3D map is also included in Figure 4 (i) for better understanding. The false coloured intensity map shows maximum hot spot strength and field distribution across the pit and base areas and of the i-NPyr nanostructures.

**3.2.1 Protein Sensing, Sweat monitoring, and Signal Stability under mechanical condition.** In order to validate our methodology and application of the SERS substrate, hemoglobin (Hb) protein in the native state was further examine in this study. The SERS spectrum (red) of $10^{-4}$ M Hb was compared with the Raman spectrum (blue) of lyophilized Hb protein powder and Raman spectrum (black) of $10^{-4}$ M Hb protein coated on Si substrate in Figure 5 (a). The distinct marker bands at 1374, 1585 and 1640 cm$^{-1}$, related to heme groups porphyrin ring modes are in good agreements [36]. The subtle peaks at 1346 cm$^{-1}$ and 1374 cm$^{-1}$ represents the ferrous and ferric oxidation states marker bands respectively [37]. Though 532 nm laser source was used to conduct the SERS study but the identical marker modes between Hb Raman spectrum and SERS spectra indicate the retention of Hb native states.

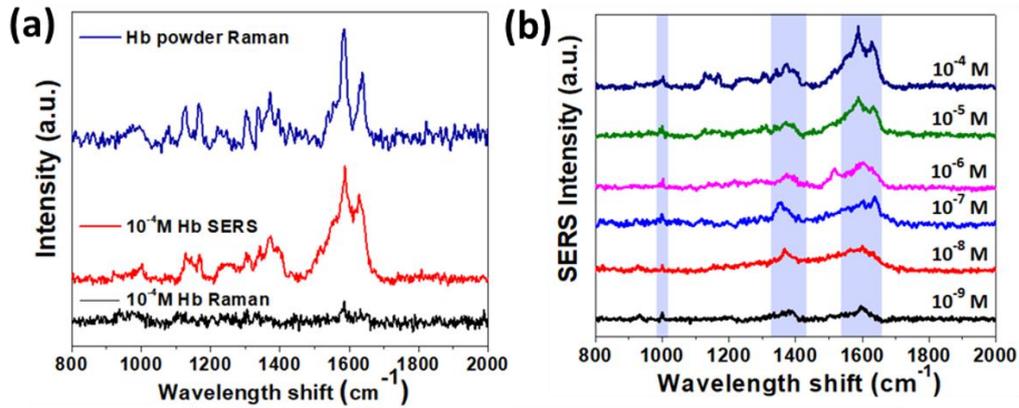
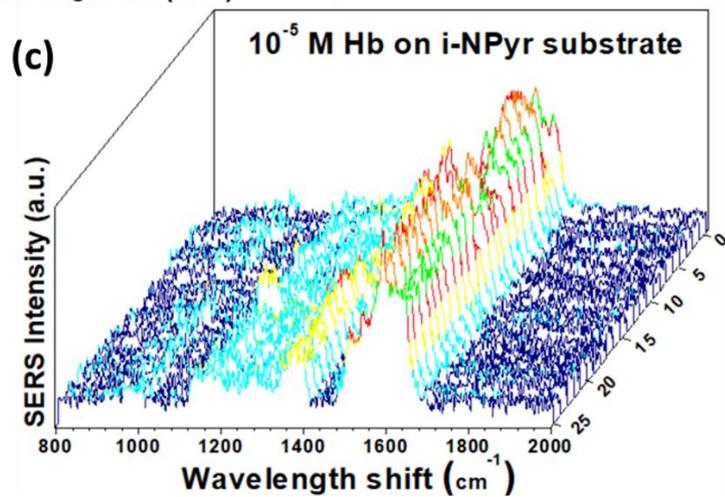
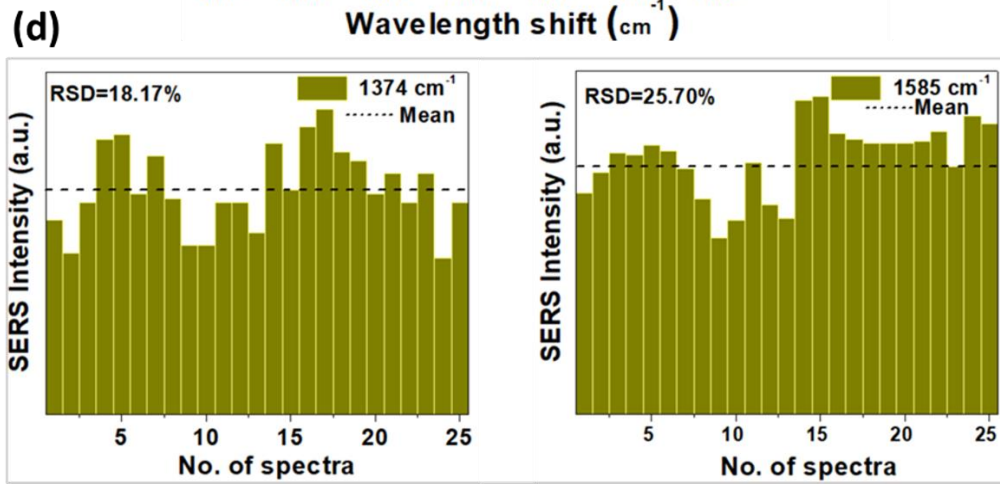
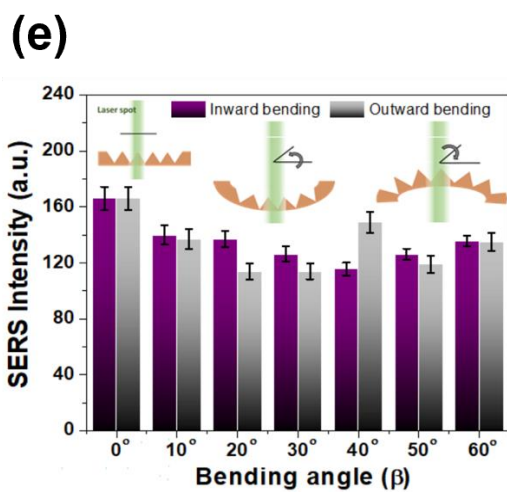
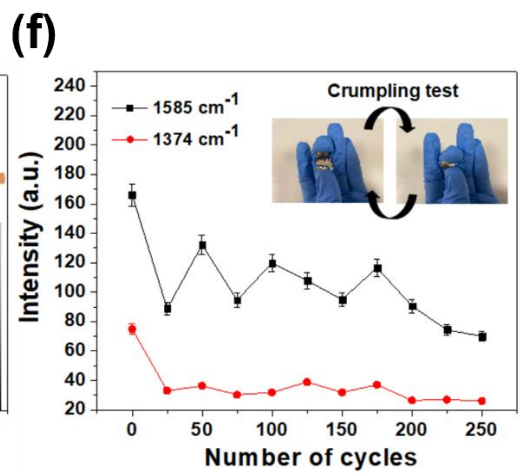

**Figure 5.** SERS detection of hemoglobin (Hb) protein on Ag/Au i-NPyr substrate. **(a)** The Raman spectra of the powder and $10^{-4}$ M Hb protein solution on Si substrate, and SERS spectrum of $10^{-4}$ M Hb protein. **(b)** Concentration-dependent SERS spectra of Hb protein under 532 laser excitation. The concentration of the protein molecules was decreased up to $10^{-9}$ M. **(c)** 3D stacking plot of SERS spectra of $10^{-5}$ M Hb protein obtained from Ag/Au nano i-NPyr substrate. **(d)** The histograms of Hb characteristic peaks intensities at 1374 cm$^{-1}$ (RSD of 18.17%) and 1585 cm$^{-1}$ (RSD of 25.70%) across 25 SERS spectra that are shown in (c). (e) The characteristic Raman peak of Hb protein of Ag/Au i-NPyr substrate under inward and outward bending over a range of excitation angle from 0° to 60° (f) characteristic Raman peak intensity during 250 cycles of crumpling test. The inset image shows the SERS sensor was held between fingers in opened and closed positions.

All assigned Hb Raman peaks including the SERS peaks at 750, 1003, 1168, 1346, and 1640 cm$^{-1}$ was listed in Table S1, in the supporting information section S5 [37][38][39][40][41]. As presented in Figure 5 (b), the SERS spectra of Hb on i-NPyr substrate in the concentration range of $10^{-4}$ to $10^{-9}$ M depicts the ability of the proposed substrate to give detectable SERS intensity at nano molar concentrations. Further, Figure 5 (c), displays SERS spectra of $10^{-5}$ M Hb proteins obtained from 25 different locations of a single i-NPyr substrate showing an outstanding SERS based detection reliability. Figure 5 (d) represents the histograms of peak intensities at 1374 cm$^{-1}$ and 1585 cm$^{-1}$ across the 25 SERS spectra with relative standard deviation (RSD) as 18.17 % and 25.70 %, respectively. Hence, the stability and reproducibility of SERS spectra proves the consistent SERS enhancements throughout the proposed substrate, which satisfies the goal of the entire fabrication process.

The signal stability of the flexible Ag/Au i-NPyr substrate under mechanical conditions, including bending and crumpling, was investigated. Figure 5 (e) shows the SERS spectra of $10^{-4}$ M Hb when the Ag/Au i-NPyr substrate was bent by an angle (β) varying from 0° to 60°. In both inward and outward bending modes, the SERS intensity for the peak 1585 cm$^{-1}$ was kept almost in the same range with no significant variation. It can be observed that the signal strength slightly decreased in bending modes as compared to the normal mode sensing. For inward and outward bending test, the signals were recorded at every 10° bending and observed variation in intensity over inward and outward bending angle ranging from 0° to 60° were shown as 14.76%, 18% and the RSD is 10.92%, 13.61% respectively. Further, the crumpling test (Figure 5 (f)) was conducted by adhering the substrate to a hand glove and then by closing and opening the hand 250 times. In this test the RSD was calculated to be 25.12% and 36.38% for 1585 cm$^{-1}$ and 1374 cm$^{-1}$ respectively. No appreciable change was observed in the Raman signal intensity even after the 250 crumpling cycle. The stability and homogeneity of SERS signal intensity under different mechanical conditions would be of great interest in wearable diagnosis.

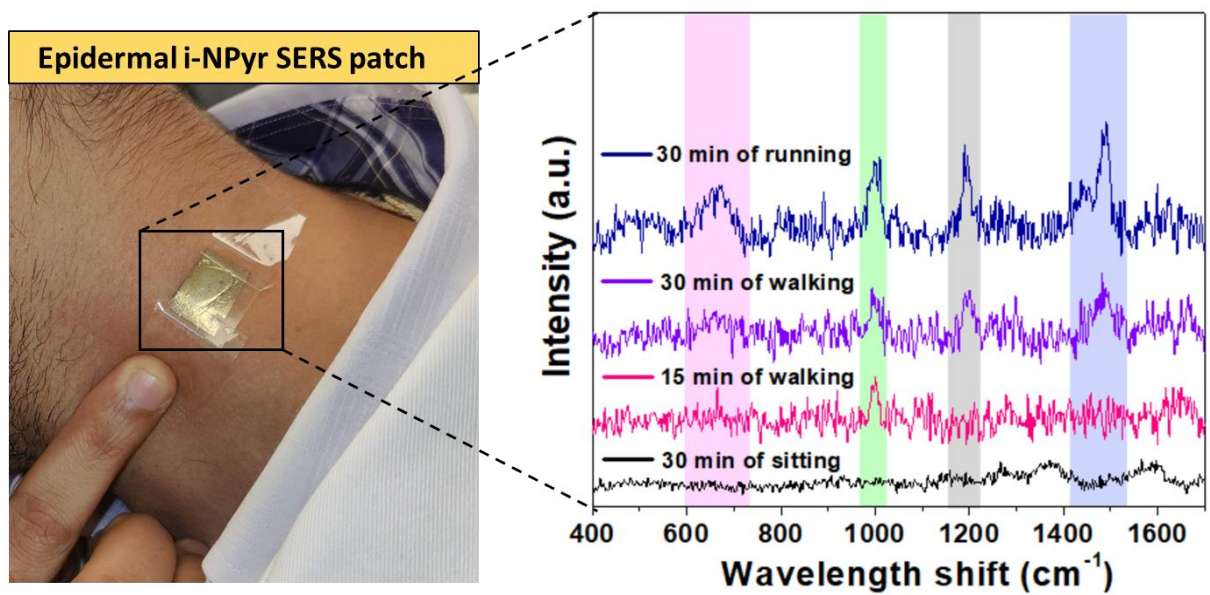

Figure 6. The average SERS spectra of dry sweat sample on i-NPyr substrate, the inset is the photograph of wearable i-NPyr sensor adhered to volunteer's neck.

Furthermore, as a proof-of concept in real application and the ability as a wearable sensor of the fabricated SERS biosensor, the flexible i-NPyr substrate was sticked as a wearable device on neck of a volunteer to analyse the human sweat sample, under different physical conditions including sitting, walking, and running (Figure 6). The sweat is most readily available among all biofluids and the monitoring of several metabolites, electrolytes and trace molecules present in the sweat sample is important to understand physiological health status. For the 30 min "sitting" state, no characteristic peaks of sweat were found. But after 15-30 min walking (85-95 m min$^{-1}$), the sweat secretion rate was increased, and in the "walking" state Raman spectra, the signature sweat Raman bands at 662, 886, 1003, 1009, 1410-1420, and 1453 cm$^{-1}$ were observed with time laps. The 1003 cm$^{-1}$ peak corresponds to the aromatic ring vibrations (CN) for urea, amino acid, and proteins. The bands in the range 1410-1420 cm$^{-1}$ and 1453 cm$^{-1}$ contain mainly urea (CN) and lactic acid (C−CH$_3$) [42]. All the assigned sweat Raman bands are listed in Table S2 in the electronic supplementary information section S5. After 15 min of the "running" phase, the concentration of sweat's constituent molecules increased more rapidly than in the walking phase. It exhibited precise SERS spectra dominated by lactic acid, urea, proteins and single amino acids. The sweat sample is heterogeneous, and its biochemical composition varies notably with the donor. Hence no Raman spectrum could adequately consider as sweat traces. These results firmly support the practical usability of the fabricated flexible i-NPyr substrate as a wearable SERS biosensor.

### 3.3 Theoretical Simulation

Although our experimental finding has confirms the better SERS performance of the fabricated i-NPyr substrates over u-NPyr substrate. However, for better understanding as well as identify the perfect SERS behaviours of the proposed SERS substrate, we have analysed induced local electromagnetic field associated with nanostructures using commercial COMSOL software. In the simulation, basically two types of nano systems i.e. Ag/Au coated polymer i-NPyr and u-NPyr were designed and simulated, and the details of the used modelling parameters, geometries and physics are included in the supplementary information section S6.

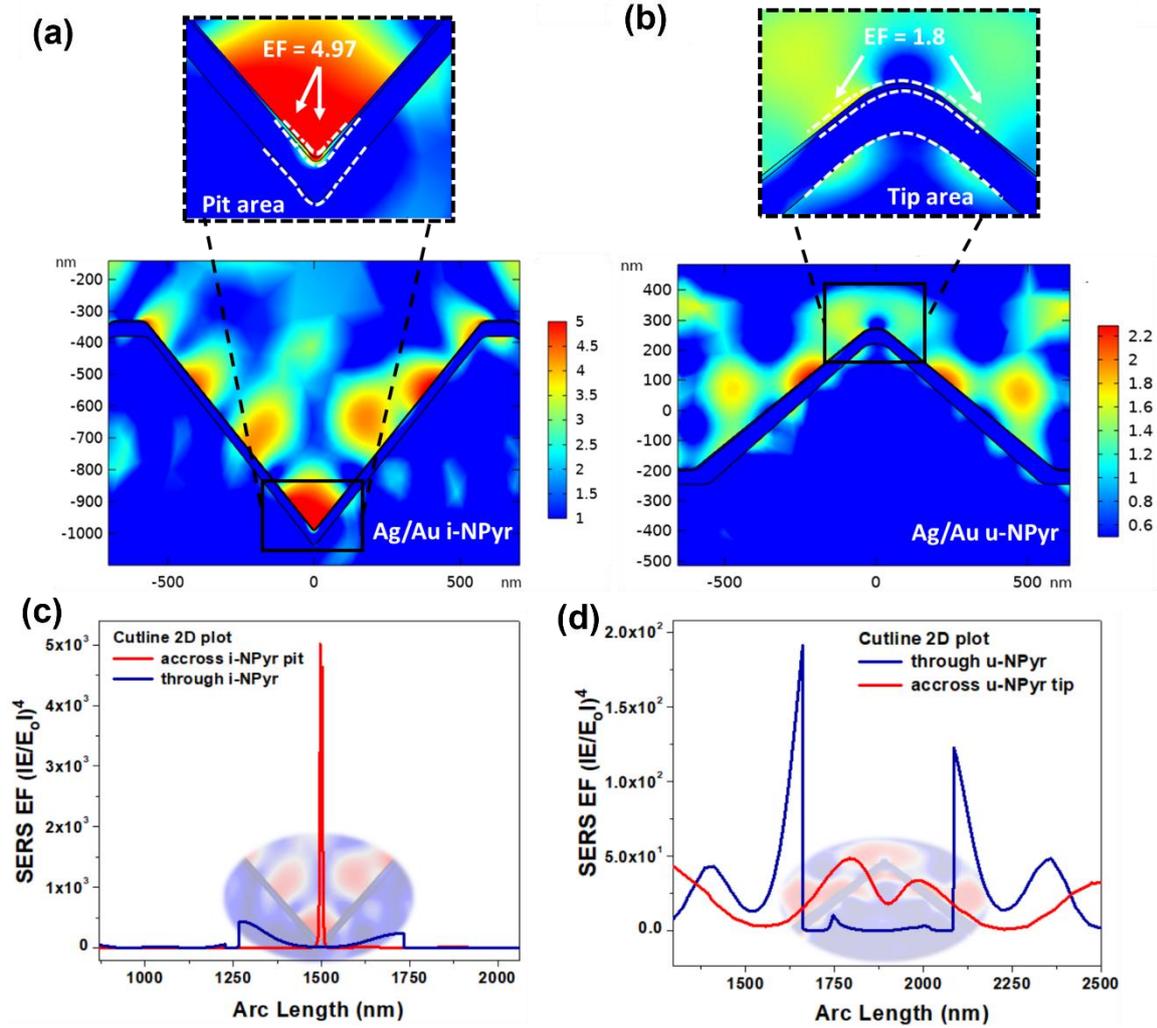

**Figure 7** Visualizations of the computed normalised electric field of **(a)** Ag/Au coated i-NPyr substrate, **(b)** Ag/Au coated u-NPyr substrate with 532 nm excitation wavelength. To ease visualization, **(a-b)** are in logarithmic scale (base 10) with zoomed field map of the i-NPyr and u-NPyr tip portions. The enhanced electric field strength is plotted as a function of distance (arc length) across the tip and middle section of **(c)** i-NPyr, and **(d)** the u-NPyr.

To accurately model the field distribution of fabricated nanostructure we have used geometrical dimensions obtained from the experimental characterization. Also, to avoid singularities of solutions, we have applied appropriate fillet condition across nano tips to get a tapered curvature. Figure 7 (a) shows the simulated electric field profiles for Ag/Au nano i-NPyr, have an intense confinement of electric field across tip areas with a maximum EF of 4.97 (on $\log_{10}$ scale) can be seen. A clear variation in field strength along the pit walls can be also visualised in the field map. However, in case of u-NPyr structure, the enhanced field is mostly confined along pyramid side walls and the observed EF across its tip area is 1.8, which is shown in Figure 7 (b). The EM SERS enhancement across the tip and the side walls of both i-NPyr and u-NPyr structures are shown in Figure 7 (c) and (d), respectively. From the simulated field profiles, it is clear that the EM enhancement is maximum in nano i-NPyr structures over u-NPyr which agrees with our experimental observations.

The maximum field distribution in the i-NPyr structure purely due to propagating surface plasmons along the pit side walls and localized plasmon across tip region [43]. There may be chances of enhancement due to the diffraction of the light beam from the i-NPyr geometry due to resonant micrometric cavity effect [44]. Further, to investigate the role of Ag/Au bimetallic layer on the SERS performance, a 50 nm Au coated i-NPyr was also simulated at an excitation wavelength of 532 nm and the resulted electric field profile is depicted in Figure S8. The observed EM enhancement in Ag/Au i-NPyr was found to be large as compared to Au i-NPyr. Earlier reports have suggested that the availability of higher number of electrons in the bimetallic system interact with incident light and produces strong plasmon induce electric field [45]. The Ag/Au i-NPyr SERS performance was also checked under 785 nm wavelength of excitation. A maximum EF of 3.12 was observed across the tip region and the corresponding field profile is shown in Figure S9.

## 4. Discussion

Silicon klarite SERS substrate is one of the standard SERS substrate and commercially produced by D3 Technologies Ltd [46]. But for regular practical applications, we need to consider both performance and cost. Nanoscale structures through nano-replica method can reduce the fabrication cost significantly [47]. Also, large-scale sample production can be done outside the cleanroom. In this study, the recipe for fabrication of the commercial grade Si i-NPyr substrate was developed. Later, using it as a master, the mass fabrication of flexible polymer u-NPyr and i-NPyr substrate were done through multi-step replica molding. Unlike previous studies, two designs, u-NPyr and i-NPyr with bimetallic coatings were explored, in order to uncover potential optimization opportunities. To achieve optimal SERS enhancement and prevent oxidation of the Ag layer, 5 nm of Au was coated as a protective layer on top of the 45 nm Ag. The metallization can further prevent the polymer substrate damage or burn up under laser excitation. Additionally, the substrate flexibility has the potential to be applied to irregularly and soft (human skin or textile fabrics etc.) surfaces seamlessly without altering their functionalities [48]. After highly precise structural batch-to-batch reproducibility, even after multiple replications (confirmed with FESEM and AFM analysis), the price per unit substrate was estimated to be less than 1 $, which could be fabricated in large scale and used as cost effective SERS platform.

Both designs were analysed experimentally and theoretically to understand their SERS enhancement ability of the nanostructure. The selected probe molecule Rh6G showed higher SERS EF on Ag/Au nano i-NPyr substrates over u-NPyr substrate. An enhancement factor of the order $10^6$ was achieved at $10^{-10}$ M concentration of the probed molecule. It is worth noted that with i-NPr, the Rh6G molecule was detected up to $10^{-12}$ M concentration. At this concentration, the ambiguity in accuracy with calculating $N_{SERS}$ (Mentioned in section S4 of supplementary information) is higher due to inability in tracing the casted drop accurately, so we adhered with $10^{-10}$ M concentration for EF calculation. Further, to support the experimental results, the EM field simulation results show large electric field confinement on the side walls of the i-NPyr and the maximum field confinement is at tip areas. Other than EF, i-NPyr substrates shows a minimum detection limit of Rh6G at picomolar concentrations with good linearity ($R^2$=0.96), which confirms further the signal reproducibility of the i-NPyr substrate. To optimize SERS for biomedical applications, we have also theoretically calculated the wavelength dependant EF under both 532 and 785 nm excitations, and the i-NPyr substrate exhibits strong electric field confinement for both excitations with relatively better results under 532 laser excitations. The deformation resulting from bending or stretching of the substrate may change the optical property, indicating a tunable plasmon resonance and can be operable under different wavelengths of laser excitations. Since field confinement can be seen for visible (532 nm) and infrared (785 nm) wavelengths, the potential of designed i-NPyr can further be explored as a broadband SERS substrate [49]. We further utilized the excellent SERS properties of the i-NPr substrate for the detection of hemoglobin protein. The large-scale structural uniformity of fabricated i-NPyr substrate gives high SERS signal reproducibility over the whole substrate with an intensity deviation of <18% measured from 25 randomly chosen points. Furthermore, the detectable concentration for Hb molecules on i-NPr substrate was down to 1 nM, which is much lowered as compared to previously reported values [50][51][52]. It is worth mentioning that most of the commercially available or chemically etched randomly arranged Si i-NPyr or u-NPyr were mainly employed for chemical trace detections.  But the use of the fabricated Ag/Au coated polymer i-NPyr SERS substrate for protein sensing has opened a new application possibility for klarite substrates as a label free biosensor. Additionally, the substrate flexibility has the potential to be applied to irregular and soft (human skin or textile fabrics etc.) surfaces seamlessly without altering their functionalities [53]. The protein sensing ability of i-NPyr substrate under different physical and mechanical conditions were examined and the signalling stability shows its potential towards portable wearable sensor.  This study assessed feasibility of i-NPyr substrate as a real wearable sensor in a sweat sample of middle-aged male volunteer. Due to intrinsic flexibility nature the i-NPyr substrates can be used as transdermal patches to fix on the skin.

**5. Conclusions**

In conclusion, we have presented a novel and efficient method to fabricate flexible Ag/Au coated polymer i-NPyr and u-NPyr substrates. Combining the good SERS activity of

Ag and chemically inertness of Au and large field confinement of pyramidal pits, the fabricated SERS substrates exhibit good stability, sensitivity, reproducibility, and homogeneity. The SERS results using Rh6G as probe molecule indicate the fabricated i-NPyr substrate is superior compared to u-NPyr substrate. A sensitive detection of hemoglobin protein was achieved using Ag/Au coated polymer i-NPyr substrate under 532 nm laser excitation. The SERS results not only confirmed the ultrasensitive detection of protein molecules at nanomolar concentrations but also profoundly resolved their structural details. Further, we have demonstrated the idea of wearable SERS sensor using our fabricated i-NPyr substrate in an epidermal patch form by detecting protein as well as monitoring sweat samples under broad range of excitation angle and physical conditions. The EM SERS behaviour are also demonstrated theoretically using COMSOL simulation and the obtained results revels EM field confinements across nano tip and side walls of the nanostructures. Also, the simulation results were in good agreement with experimental SERS enhancement values. This approach provides us a capital idea to accelerate the development of low coast label-free flexible SERS nano biosensor for the point-of-care diagnostics.

**References:**

1    Choi S, Lee H, Ghaffari R, Hyeon T, Kim DH. Recent Advances in Flexible and Stretchable Bio-Electronic Devices Integrated with Nanomaterials. *Adv. Mater.* 28(22) (2016).

2    Wang X, Liu Z, Zhang T. Flexible Sensing Electronics for Wearable/Attachable Health Monitoring, (2017).

3    Liu Y, Pharr M, Salvatore GA. Lab-on-Skin: A Review of Flexible and Stretchable Electronics for Wearable Health Monitoring, (2017).

4    Bandodkar AJ, Jeerapan I, Wang J. Wearable Chemical Sensors: Present Challenges

and Future Prospects. *ACS Sensors* 1(5) (2016).

5    Kim J, Campbell AS, de Ávila BEF, Wang J. Wearable biosensors for healthcare monitoring, (2019).

6    Gao F, Liu C, Zhang L *et al.* Wearable and flexible electrochemical sensors for sweat analysis: a review. *Microsystems Nanoeng.* 9(1) (2023).

7    Kim J, Kumar R, Bandodkar AJ, Wang J. Advanced Materials for Printed Wearable Electrochemical Devices: A Review, (2017).

8    Liu YL, Qin Y, Jin ZH *et al.* A Stretchable Electrochemical Sensor for Inducing and Monitoring Cell Mechanotransduction in Real Time. *Angew. Chemie - Int. Ed.* 56(32) (2017).

9    Han XX, Zhao B, Ozaki Y. Surface-enhanced Raman scattering for protein detection, (2009).

10    Feliu N, Hassan M, Garcia Rico E, Cui D, Parak W, Alvarez-Puebla R. SERS Quantification and Characterization of Proteins and Other Biomolecules. *Langmuir* 33(38) (2017).

11    Drescher D, Büchner T, McNaughton D, Kneipp J. SERS reveals the specific interaction of silver and gold nanoparticles with hemoglobin and red blood cell components. *Phys. Chem. Chem. Phys.* 15(15) (2013).

12    Muhammad P, Hanif S, Yan J *et al.* Correction: SERS-based nanostrategy for rapid anemia diagnosis. *Nanoscale* 13(37) (2021).

13    González-Solís JL, Martínez-Espinosa JC, Salgado-Román JM, Palomares-Anda P. Monitoring of chemotherapy leukemia treatment using Raman spectroscopy and principal component analysis. *Lasers Med. Sci.* 29(3) (2014).

14  Feng S, Chen R, Lin J *et al.* Nasopharyngeal cancer detection based on blood plasma surface-enhanced Raman spectroscopy and multivariate analysis. *Biosens. Bioelectron.* 25(11) (2010).

15  Berger AJ, Itzkan I, Feld MS. Feasibility of measuring blood glucose concentration by near-infrared Raman spectroscopy. *Spectrochim. Acta - Part A Mol. Spectrosc.* 53(2) (1997).

16  Song CY, Yang YJ, Yang BY, Sun YZ, Zhao YP, Wang LH. An ultrasensitive SERS sensor for simultaneous detection of multiple cancer-related miRNAs. *Nanoscale* 8(39) (2016).

17  Koh EH, Lee WC, Choi YJ *et al.* A wearable surface-enhanced raman scattering sensor for label-free molecular detection. *ACS Appl. Mater. Interfaces* 13(2) (2021).

18  Wu J, Zhou X, Li P *et al.* Ultrasensitive and Simultaneous SERS Detection of Multiplex MicroRNA Using Fractal Gold Nanotags for Early Diagnosis and Prognosis of Hepatocellular Carcinoma. *Anal. Chem.* 93(25) (2021).

19  Fan M, Andrade GFS, Brolo AG. A review on the fabrication of substrates for surface enhanced Raman spectroscopy and their applications in analytical chemistry, (2011).

20  Wu J, Fang J, Yang X, Wang C. Large-scale flexible metal-covered polymer nanopillar arrays as highly uniform and reproducible SERS substrates for trace analysis. *Nanotechnology* 29(46) (2018).

21  Liao PF, Wokaun A. Lightning rod effect in surface enhanced Raman scattering, (1982).

22  Fukuoka N, Tanabe K. Lightning-rod effect of plasmonic field enhancement on hydrogen-absorbing transition metals. *Nanomaterials* 9(9) (2019).

23  Tahir MA, Zhang X, Cheng H *et al.* Klarite as a label-free SERS-based assay: A promising approach for atmospheric bioaerosol detection. *Analyst* 145(1) (2020).

24  Almaviva S, Palucci A, Botti S, Puiu A, Rufoloni A. Validation of a Miniaturized Spectrometer for Trace Detection of Explosives by Surface-Enhanced Raman Spectroscopy. *Challenges* 7(2) (2016).

25  Palermo G, Rippa M, Conti Y *et al.* Plasmonic Metasurfaces Based on Pyramidal Nanoholes for High-Efficiency SERS Biosensing. *ACS Appl. Mater. Interfaces* (2021).

26  Xu G, Cheng H, Jones R *et al.* Surface-Enhanced Raman Spectroscopy Facilitates the Detection of Microplastics <1 μm in the Environment. *Environ. Sci. Technol.* 54(24) (2020).

27  Wang Y, Lu N, Wang W *et al.* Highly effective and reproducible surface-enhanced Raman scattering substrates based on Ag pyramidal arrays. *Nano Res.* 6(3) (2013).

28  Zhang J, Yan Y, Miao P, Cai J. Fabrication of gold-coated PDMS surfaces with arrayed triangular micro/nanopyramids for use as SERS substrates. *Beilstein J. Nanotechnol.* 8(1) (2017).

29  Oo SZ, Chen RY, Siitonen S *et al.* Disposable plasmonic plastic SERS sensor. *Opt. Express* 21(15) (2013).

30  Das A, Moirangthem RS. Plasmonic Nanoprobes for SERS-Based Theranostics Applications. In: *Recent Advances in Plasmonic Probes* (*Volume 33*). Biswas R, Mazumder N (Ed.), Springer, Cham, 223–244 (2022).

31  Bodelón G, Montes-García V, Pérez-Juste J, Pastoriza-Santos I. Surface-enhanced Raman scattering spectroscopy for label-free analysis of P. aeruginosa Quorum Sensing, (2018).

32  Liu M, Guyot-Sionnest P. Synthesis and optical characterization of Au/Ag core/shell nanorods. *J. Phys. Chem. B* 108(19) (2004).

33  Das, Anindita Pant, Udit Cao C, Moirangthem RS, Kamble HB. Fabrication of plasmonic nanopyramidal array as flexible SERS substrate for biosensing application. *Nano Res.* (2022).

34  Kern W. Evolution of silicon wafer cleaning technology. Presented at: *Proceedings - The Electrochemical Society*. 1990.

35  Lindberg FW, Norrby M, Rahman MA *et al.* Controlled Surface Silanization for Actin-Myosin Based Nanodevices and Biocompatibility of New Polymer Resists. *Langmuir* 34(30) (2018).

36  Xu H, Bjerneld EJ, Käll M, Börjesson L. Spectroscopy of single hemoglobin molecules by surface enhanced raman scattering. *Phys. Rev. Lett.* 83(21) (1999).

37  Kalaivani G, Sivanesan A, Kannan A, Venkata Narayanan NS, Kaminska A, Sevvel R. Plasmon-tuned silver colloids for SERRS analysis of methemoglobin with preserved nativity. *Langmuir* 28(40) (2012).

38  Kang Y, Si M, Liu R, Qiao S. Surface-enhanced Raman scattering (SERS) spectra of hemoglobin on nano silver film prepared by electrolysis method. *J. Raman Spectrosc.* 41(6) (2010).

39  Hu S, Smith KM, Spiro TG. Assignment of protoheme Resonance Raman spectrum by heme labeling in myoglobin. *J. Am. Chem. Soc.* 118(50) (1996).

40  *Vibrational Spectroscopy in Protein Research*. (2020).

41  Casella M, Lucotti A, Tommasini M *et al.* Raman and SERS recognition of β-carotene and haemoglobin fingerprints in human whole blood. *Spectrochim. Acta - Part A Mol.*

*Biomol. Spectrosc.* 79(5) (2011).

42   Sikirzhytski V, Sikirzhytskaya A, Lednev IK. Multidimensional Raman spectroscopic signature of sweat and its potential application to forensic body fluid identification. *Anal. Chim. Acta* 718 (2012).

43   Perney NMB, Baumberg JJ, Zoorob ME, Charlton MDB, Netti CM. Tuning localized plasmons in nanostructured substrates for surface-enhanced Raman scattering applications. Presented at: *Optics InfoBase Conference Papers*. 2006.

44   Degioanni S, Jurdyc AM, Cheap A *et al.* Surface-enhanced Raman scattering of amorphous silica gel adsorbed on gold substrates for optical fiber sensors. *J. Appl. Phys.* 118(15) (2015).

45   Mulyanti B, Wulandari C, Mohamad NR, Rifaldi E, Hasanah L, Menon PS. Bimetallic ag/au thin films in kretschmann-based surface plasmon resonance sensor for glucose detection. *Optoelectron. Adv. Mater. Rapid Commun.* 14(11–12) (2020).

46   Hankus ME, Stratis-Cullum DN, Pellegrino PM. Surface enhanced Raman scattering (SERS)-based next generation commercially available substrate: physical characterization and biological application. Presented at: *Biosensing and Nanomedicine IV*. 2011.

47   Choi CJ, Xu Z, Wu HY, Liu GL, Cunningham BT. Surface-enhanced Raman nanodomes. *Nanotechnology* 21(41) (2010).

48   Fu HY, Lang XY, Hou C *et al.* Nanoporous Au/SnO/Ag heterogeneous films for ultrahigh and uniform surface-enhanced Raman scattering. *J. Mater. Chem. C* 2(35) (2014).

49   Mao P, Liu C, Chen Q, Han M, Maier SA, Zhang S. Broadband SERS detection with disordered plasmonic hybrid aggregates. *Nanoscale* 12(1) (2020).

50   Raju NRC. Hemoglobin detection on AgO surface enhanced Raman scattering (SERS)-substrates. *Mater. Lett.* 130 (2014).

51   Pal AK, Chandra GK, Umapathy S, Bharathi Mohan D. Ultra-sensitive, reusable, and superhydrophobic Ag/ZnO/Ag 3D hybrid surface enhanced Raman scattering substrate for hemoglobin detection. *J. Appl. Phys.* 127(16) (2020).

52   Ding Q, Kang Y, Li W *et al.* Bioinspired Brochosomes as Broadband and Omnidirectional Surface-Enhanced Raman Scattering Substrates. *J. Phys. Chem. Lett.* 10(21) (2019).

53   Costa JC, Spina F, Lugoda P, Garcia-Garcia L, Roggen D, Münzenrieder N. Flexible Sensors—From Materials to Applications. *Technologies* 7(2) (2019).